\documentclass[aps,amsmath,amssymb,twocolumn,prd,floatfix,superscriptaddress,10pt,showpacs,showkeys]{revtex4-1}

\usepackage{graphicx}
\usepackage{amssymb}
\usepackage{amsmath}
\usepackage{dcolumn}% Align table columns on decimal point
\usepackage{docs}%
\usepackage{bm}% bold math
\usepackage{color}% annotate MS with colored text
\usepackage{ulem}
\def\lsim{\mathrel{\rlap{\lower4pt\hbox{\hskip1pt$\sim$}}
    \raise1pt\hbox{$<$}}}               
\def\gsim{\mathrel{\rlap{\lower4pt\hbox{\hskip1pt$\sim$}}
    \raise1pt\hbox{$>$}}} 

\begin{document}

\title{Freeze-out conditions from strangeness observables at RHIC}

\author{Marcus Bluhm}
\affiliation{Institute of Theoretical Physics, University of Wroclaw, PL-50-204 Wroclaw, Poland}
\email{marcus.bluhm@uwr.edu.pl}
\author{Marlene Nahrgang}
\affiliation{SUBATECH UMR 6457 (IMT Atlantique, Universit\'e de Nantes, \\IN2P3/CNRS), 4 rue Alfred Kastler, 44307 Nantes, France}
\email{marlene.nahrgang@subatech.in2p3.fr}

\keywords{chemical freeze-out, net-Kaon fluctuations, strangeness production, QCD phase diagram}
\pacs{12.38.Mh, 25.75.Nq}

\begin{abstract}
We determine chemical freeze-out conditions from strangeness observables measured at RHIC beam energies. Based on a combined 
analysis of lowest-order net-Kaon fluctuations and strange anti-baryon over baryon yield ratios we obtain visibly enhanced 
freeze-out conditions at high beam energies compared to previous studies which analyzed net-proton and net-charge fluctuations. Our 
findings are in qualitative agreement with the recent study~\cite{Bellwied:2018tkc} which utilizes the net-Kaon fluctuation data in 
combination with information from lattice QCD. Our complimentary approach shows that also strange hadron yield ratios are described 
by such enhanced freeze-out conditions.
\end{abstract}

\maketitle

%%%%%%%%%%%%%%%%%%%%%%%%%%%%%%%%%%%%%%%%%%%%%%%%%%%%%%%%%%%%%%%%%%%%%%%%%%%
\section{Introduction \label{sec:1}}
%%%%%%%%%%%%%%%%%%%%%%%%%%%%%%%%%%%%%%%%%%%%%%%%%%%%%%%%%%%%%%%%%%%%%%%%%%%

High-energy heavy-ion collision experiments at various beam energies $\sqrt{s}$ have enriched our understanding of the properties 
and phases of strongly interacting matter. The transient creation of a color-deconfined state in the laboratories was one of the 
major scientific successes in the last two decades. The deconfinement transition is an analytic crossover for vanishing baryon 
chemical potential $\mu_B$~\cite{Aoki:2006we}, where the transition region 
$T_c=(154\pm 9)$~MeV~\cite{Borsanyi:2010bp,Bazavov:2011nk} is rather broad in temperature $T$. While these information base on 
first-principle lattice QCD calculations, another fascinating landmark in the phase diagram, the QCD critical point, has not yet 
been discovered with this method despite being predicted by various approaches~\cite{Berges:1998rc,Halasz:1998qr}. 

Information about the properties of hot and dense QCD matter can be inferred indirectly from the measured particle spectra and 
their event-by-event fluctuations. The success of statistical hadronization models in describing the particle production in 
heavy-ion collisions ranging from AGS to LHC beam 
energies~\cite{Cleymans:2005xv,Becattini:2005xt,Andronic:2008gu,Andronic:2011yq} led to the conclusion that the produced hadronic 
matter originates from a source in or near thermal and chemical equilibrium. A common freeze-out curve~\cite{Cleymans:2005xv} in 
the phase diagram could be drawn, highlighting the thermal conditions ($T, \mu_B$) at chemical freeze-out where the hadrochemistry 
is fixed. 

Higher-order moments of the event-by-event particle multiplicity distributions provide an additional excellent measure to 
characterize the matter properties and to reveal the phase structure in QCD. While it is debated whether fluctuations originate 
from an equilibrated hadronic medium~\cite{Kitazawa:2013bta,Sakaida:2014pya} it is to a first extent reasonable to assume, that if 
this is true for the means also the lowest-order fluctuations, i.e.~the variances, should be describable within statistical models. 
First studies~\cite{Alba:2014eba,Borsanyi:2014ewa}, utilizing experimental data on net-proton and net-electric charge fluctuations 
to determine the freeze-out conditions, found at large $\sqrt{s}$ significantly reduced freeze-out parameters compared 
to~\cite{Cleymans:2005xv}. In the work~\cite{Alba:2014eba}, the analysis was quantitatively driven by the contributions from 
protons and anti-protons as well as charged pions. 

In a recent study~\cite{Bellwied:2018tkc}, the experimental data~\cite{Adamczyk:2017wsl} on lowest-order fluctuations in the 
net-Kaon number $N_{K^+}-N_{K^-}$ as function of $\sqrt{s}$ were used to extract freeze-out conditions. For a unique determination, 
information from lattice QCD for the isentropic trajectories~\cite{Gunther:2016vcp} running through the freeze-out points 
of~\cite{Alba:2014eba} was supplemented. The obtained freeze-out temperatures were reported to be substantially larger than those 
found in~\cite{Alba:2014eba}. In our work, we combine an analysis of the net-Kaon fluctuations~\cite{Adamczyk:2017wsl} and the 
yield ratios of strange anti-baryons over baryons~\cite{Adams:2006ke,Aggarwal:2010ig,Zhao:2013yza} as functions of $\sqrt{s}$. 
This complimentary approach allows us to determine whether strange baryon yields and net-Kaon fluctuations can be described with 
the same $T$ and $\mu_B$. 

Our strategy is to analyze observables that are highly sensitive to variations in the thermal parameters. As we discussed 
in~\cite{Alba:2015iva}, the ratio of variance over mean of the net-Kaon number is such an observable that varies rapidly with $T$ 
in a statistical model. The same is true for the heavier baryons while we use the lighter baryons as a baryometer in this work. 
Therefore, we consider the combination of observables studied here as optimized for determining freeze-out 
conditions. One should keep in mind, nevertheless, that the net-Kaon number is not a conserved charge in QCD and as such prone 
to late stage processes like resonance decays. 

%%%%%%%%%%%%%%%%%%%%%%%%%%%%%%%%%%%%%%%%%%%%%%%%%%%%%%%%%%%%%%%%%%%%%%%%%%%
\section{Theoretical framework \label{sec:2}}
%%%%%%%%%%%%%%%%%%%%%%%%%%%%%%%%%%%%%%%%%%%%%%%%%%%%%%%%%%%%%%%%%%%%%%%%%%%

We perform our analysis of net-Kaon fluctuations and strange anti-baryon to baryon yield ratios using a Hadron Resonance Gas (HRG) 
model for a grandcanonical ensemble of non-interacting hadrons and resonances. Such a model, in which the strong interaction between 
hadrons is effectively accounted for by contributions from resonances~\cite{Venugopalan:1992hy}, was shown to reliably describe 
basic thermodynamic quantities~\cite{Karsch:2003vd,Karsch:2003zq,Tawfik:2004sw} as well as susceptibilities and their 
ratios~\cite{Borsanyi:2013hza,Mukherjee:2013lsa,Borsanyi:2011sw} from lattice QCD. The hadrons and resonances are considered to be 
point-like in our work. The pressure $P$ in this framework reads 
\begin{equation}
\label{equ:HRGpressure}
 P = \sum_i (-1)^{B_i+1} \frac{d_iT}{(2\pi)^3} \!\int \!\textrm{d}^3k \ln\left[1+(-1)^{B_i+1}z_i e^{-\epsilon_i/T}\right],
\end{equation}
where the sum runs over all particle species included in the framework. In Eq.~(\ref{equ:HRGpressure}) the particle energy reads 
$\epsilon_i=\sqrt{k^2+m_i^2}$ for momentum $k$ and particle mass $m_i$, $d_i$ is the degeneracy factor and $z_i=e^{\mu_i/T}$ is the 
fugacity. The particle chemical potential $\mu_i$ is defined as $\mu_i=B_i\mu_B+S_i\mu_S+Q_i\mu_Q$, where $\mu_X$ denotes the 
chemical potential of the conserved charge $X$ and $X_i=B_i, S_i, Q_i$ represent the quantum numbers of the conserved baryon, 
strangeness and electric charge, respectively. 

In Eq.~(\ref{equ:HRGpressure}), we use an updated version of the spectrum of hadrons and resonances in line with the recent listing 
from the Particle Data Group~\cite{Patrignani:2016xqp}. In previous studies~\cite{Alba:2017hhe,Alba:2017mqu}, the advantages of 
using such an update were discussed in detail. Moreover, in~\cite{Chatterjee:2017yhp} the influence of unconfirmed resonances in the 
2016 listing~\cite{Patrignani:2016xqp} versus only confirmed resonances on determined freeze-out parameters was investigated. 

The net-density $n_X$ of a conserved charge is given by $n_X=\sum_i X_i n_i$, where the individual particle densities $n_i$ follow 
from derivatives, $n_i=(\partial P/\partial\mu_i)$, at fixed $T$. With this, physical conditions met in a heavy-ion collision 
experiment can be implemented into the model~\cite{Karsch:2010ck} by requiring that $n_S=0$ and $n_Q=x n_B$. These account for 
net-strangeness neutrality in the fireball and an initial proton to baryon ratio at mid-rapidity which for Au$+$Au and Pb$+$Pb 
collisions is approximately $x\simeq 0.4$. As a consequence, the chemical potentials $\mu_S$ and $\mu_Q$ become functions of $T$ and 
$\mu_B$. Due to the lack of stopping at high $\sqrt{s}$, the mid-rapidity region is almost isospin symmetric. As $\mu_B$, and thus 
$n_B$, is small for high beam energies this is also approximately satisfied by the second physical condition. 

The experimentally realized phase-space coverage, which can be limited in rapidity $y$, transverse momentum $k_T$ and azimuthal 
angle $\phi$ due to the detector design and demands from the analysis, can be respected in a straightforward way. 
Following~\cite{Garg:2013ata}, kinematic acceptance cuts are implemented into the model by restricting the momentum integrals in 
Eq.~(\ref{equ:HRGpressure}) accordingly. This requires replacing the integration measure $\textrm{d}^3k$ by 
$k_T\sqrt{k_T^2+m_i^2}\cosh(y)\,\textrm{d}k_T\,\textrm{d}y\,\textrm{d}\phi$ 
and $\epsilon_i$ by $\cosh(y)\sqrt{k_T^2+m_i^2}$. It should be noted that this 
procedure does not take into account the elastic scatterings between chemical and kinetic freeze-out, which can transport individual 
particles in and out of the experimental acceptance. In order to take this correctly into account, a fully dynamical transport 
approach for the hadronic phase would need to be applied. We think, however, that only a small number of all considered particles 
are actually affected by this kind of final state effect. 

The effect of resonance decays can be implemented into the framework in an explicit way which allows us to keep correctly track of 
the strangeness transfer from mother to daughter including resonances such as e.g.~$\Xi(1690)^-$ or $N(1650)$. This method was 
developed in~\cite{Begun:2006jf,Fu:2013gga} and applied to study the impact of resonance decays on net-proton fluctuations 
without~\cite{Nahrgang:2014fza} and in the presence of a QCD critical point~\cite{Bluhm:2016byc}. 
The final particle number $N_j$ of a stable, i.e.~with respect to strong and electromagnetic decays, hadron is given by the sum 
$N_j=N_j^*+\sum_R \langle N_j^R\rangle_R$ of primordially, i.e.~directly, produced hadrons $N_j^*$ and the contributions stemming 
from resonance decays. Those produce on average over the decays $\langle N_j^R\rangle_R=N_R^*\langle n_j\rangle_R$ hadrons of type 
$j$ associated with the branching ratios $b_r^R$ of resonance $R$ via $\langle n_j\rangle_R=\sum_r b_r^R n_{j,r}^R$ for integer 
$n_{j,r}^R$. On an event-by-event basis, the numbers $N_j^*$ and $N_R^*$ fluctuate thermally. But, in addition, the actual number 
of hadrons of type $j$ originating from the decay of $R$ follows a multinomial probability distribution~\cite{Begun:2006jf}. Thus, 
fluctuations in the contributions from resonance decays are caused both by thermal fluctuations in $N_R^*$ and by the 
probabilistic character of the decay process. For the mean, the latter has no consequences and thus the mean of the final particle 
number after resonance decays is given by 
\begin{equation}
\label{equ:mean}
 M_j = \langle N_j^*\rangle_T + \sum_R \langle N_R^*\rangle_T \langle n_j\rangle_R \,.
\end{equation}
Accordingly, for the net-Kaon number, the mean is given by $M_K=M_{K^+}-M_{K^-}$. The variance of the net-Kaon number is instead 
influenced by the probabilistic nature of the decay and follows as~\cite{Begun:2006jf,Fu:2013gga} 
\begin{align}
\nonumber
 \sigma_K^2 = & \,\langle (\Delta N_{K^+}^*)^2\rangle_T + \langle (\Delta N_{K^-}^*)^2\rangle_T \\
\nonumber
 & + \sum_R \langle (\Delta N_{R}^*)^2\rangle_T \left(\langle n_{K^+}\rangle_R^2 + \langle n_{K^-}\rangle_R^2 \right) \\
\nonumber
 & - 2 \sum_R \langle (\Delta N_{R}^*)^2\rangle_T \langle n_{K^+}\rangle_R \, \langle n_{K^-}\rangle_R \\
\nonumber
 & + \sum_R \langle N_{R}^*\rangle_T \left(\langle (\Delta n_{K^+})^2\rangle_R + \langle (\Delta n_{K^-})^2\rangle_R\right) \\
\label{equ:variance}
 & - 2 \sum_R \langle N_{R}^*\rangle_T \langle \Delta n_{K^+} \Delta n_{K^-}\rangle_R \,,
\end{align}
where $\Delta N_i=N_i-\langle N_i\rangle_T$ and 
$\langle\Delta n_i\Delta n_l\rangle_R=\langle n_i n_l\rangle_R - \langle n_i\rangle_R \langle n_l\rangle_R =
\sum_r b_r^R n_{i,r}^R n_{l,r}^R - \langle n_i\rangle_R \langle n_l\rangle_R$. As apparent from Eq.~(\ref{equ:variance}), resonance 
decays introduce correlations between the number of $K^+$ and $K^-$. Neglecting the probabilistic nature of the decay processes, 
the variance reads instead 
\begin{align}
\nonumber
 \sigma_K^2 = & \,\langle (\Delta N_{K^+}^*)^2\rangle_T + \langle (\Delta N_{K^-}^*)^2\rangle_T \\
\nonumber
 & + \sum_R \langle (\Delta N_{R}^*)^2\rangle_T \left(\langle n_{K^+}\rangle_R^2 + \langle n_{K^-}\rangle_R^2 \right) \\
\label{equ:varianceNoProb}
 & - 2 \sum_R \langle (\Delta N_{R}^*)^2\rangle_T \langle n_{K^+}\rangle_R \, \langle n_{K^-}\rangle_R \,.
\end{align}
In Eqs.~(\ref{equ:mean})~-~(\ref{equ:varianceNoProb}) only the thermal averages $\langle \cdot \rangle_T$ can be obtained from the 
HRG model pressure in Eq.~(\ref{equ:HRGpressure}) as the derivatives 
\begin{align}
 \label{equ:ParticleNumber}
 \langle N_i\rangle_T = & \, VT^3\, \frac{\partial(P/T^4)}{\partial(\mu_i/T)} \equiv V n_i \,, \\
 \label{equ:ParticleVariance}
 \langle (\Delta N_i)^2\rangle_T = & \, VT^3\, \frac{\partial^2(P/T^4)}{\partial(\mu_i/T)^2} \,,
\end{align}
where $V$ is the volume which cancels in ratios. 

%%%%%%%%%%%%%%%%%%%%%%%%%%%%%%%%%%%%%%%%%%%%%%%%%%%%%%%%%%%%%%%%%%%%%%%%%%%
\section{Analysis and results \label{sec:3}}
%%%%%%%%%%%%%%%%%%%%%%%%%%%%%%%%%%%%%%%%%%%%%%%%%%%%%%%%%%%%%%%%%%%%%%%%%%%

%%%%%%%%%%%%%%%%%%%%%%%%%%%%%%%%%%%%%%%%%%%%%%%%%%%%%%%%%%%%%%%%%%%%%%%%%%%
\begin{figure}[t]
\centering
\includegraphics[width=0.48\textwidth]{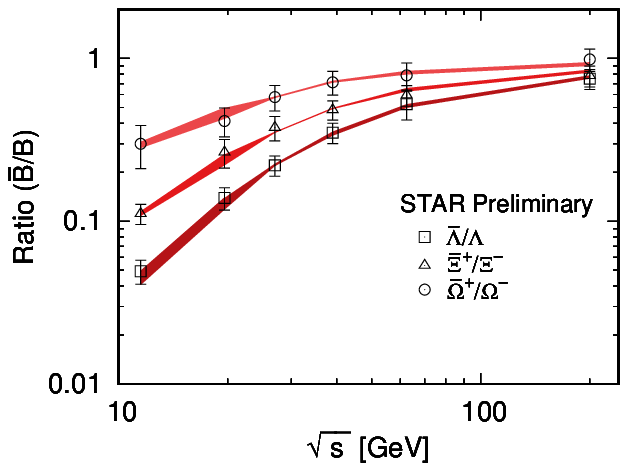}
\includegraphics[width=0.48\textwidth]{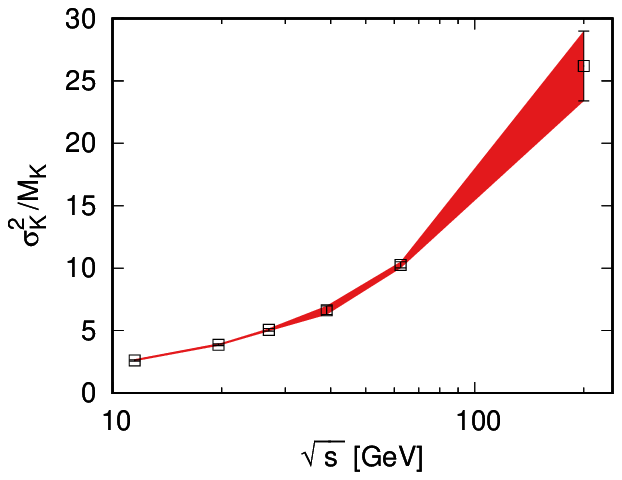}
\caption[]{\label{fig:fig1New}
(Color online) Upper panel: Ratio $\bar{B}/B$ of strange anti-baryon over baryon yields 
for $\Lambda$, $\Xi^-$ and $\Omega^-$ as a function of the beam energy $\sqrt{s}$. The published and preliminary STAR data 
(symbols) are taken from~\cite{Adams:2006ke,Aggarwal:2010ig} and~\cite{Zhao:2013yza}, respectively. The colored bands show 
the fit results for $\bar{B}/B$ within the employed HRG model for the freeze-out conditions determined in this work. Lower 
panel: Lowest-order net-Kaon fluctuation measure $\sigma_K^2/M_K$ of variance $\sigma_K^2$ over mean $M_K$ as 
function of $\sqrt{s}$. The published STAR data (squares) are taken from~\cite{Adamczyk:2017wsl}, where 
the shown error bars indicate only the dominating systematic error. The colored band depicts our fit results 
for the freeze-out conditions shown in Fig.~\ref{fig:fig2New}.}
\end{figure}
%%%%%%%%%%%%%%%%%%%%%%%%%%%%%%%%%%%%%%%%%%%%%%%%%%%%%%%%%%%%%%%%%%%%%%%%%%%
We determine the conditions for $T$ and $\mu_X$ at chemical freeze-out by applying the framework outlined above to optimally 
describe experimental data from RHIC on strangeness observables measured by the STAR Collaboration. We analyze data on yield ratios 
of strange anti-baryons over baryons, $\bar{B}/B$, as well as lowest-order net-Kaon fluctuations, $\sigma_K^2/M_K$, simultaneously. 
\begin{table}
\centering
 \begin{tabular}[t]{|l|c|c|c|c|c|c|}
  \hline
  \rule{0pt}{2ex}
  $\sqrt{s}$ & $200$ & $62.4$ & $39$ & $27$ & $19.6$ & $11.5$ \\
  \hline
  \rule{0pt}{3ex}
  $\Lambda$ & $|y|\!<\!1$ & $|y|\!<\!1$ & $|y|\!<\!0.5$ & $|y|\!<\!0.5$ & $|y|\!<\!0.5$ & $|y|\!<\!0.5$ \\
  \hline
  \rule{0pt}{3ex}
  $\Xi$ & $|y|\!<\!0.75$ & $|y|\!<\!1$ & $|y|\!<\!0.5$ & $|y|\!<\!0.5$ & $|y|\!<\!0.5$ & $|y|\!<\!0.5$ \\
  \hline
  \rule{0pt}{3ex}
  $\Omega$ & $|y|\!<\!1$ & $|y|\!<\!1$ & $|y|\!<\!0.5$ & $|y|\!<\!0.5$ & $|y|\!<\!0.5$ & $|y|\!<\!0.5$ \\
  \hline
 \end{tabular}
 \caption[]{\label{tab:tab1} Summary of considered rapidity windows in the analysis of $\bar{B}/B$-ratios. 
 The beam energy $\sqrt{s}$ is given in GeV.}
\end{table}

For the $\bar{B}/B$-ratios we study the published results in~\cite{Adams:2006ke,Aggarwal:2010ig} for $\sqrt{s}=200$ and $62.4$~GeV, 
and for smaller $\sqrt{s}$ the preliminary results from the RHIC Beam Energy Scan reported in~\cite{Zhao:2013yza}. The ratios for 
most central collisions, shown as functions of $\sqrt{s}$ in Fig.~\ref{fig:fig1New} (upper panel) for $\Lambda$ (squares), $\Xi^-$ 
(triangles) and $\Omega^-$ (circles), were determined from $\phi$- and $k_T$-integrated yields in a given rapidity window around 
mid-rapidity. In line with the experimental set-up, we consider the $y$-ranges summarized in Tab.~\ref{tab:tab1}. 

For the net-Kaon fluctuations we take the results recently reported in~\cite{Adamczyk:2017wsl} which are corrected for finite 
detector efficiency and the centrality bin width effect. The data for most central collisions are shown by the squares in 
Fig.~\ref{fig:fig1New} (lower panel) where we include only the dominating systematic error bars in the plot. For $\sigma_K^2/M_K$, 
the experimental kinematic acceptance limitations are $0.2$~GeV/c~$\leq k_T \leq 1.6$~GeV/c and $-0.5\leq y\leq 0.5$ with full 
azimuthal coverage. 

In the analysis of the data, the contributions from resonance decays play an important role. For example, for the net-Kaon 
fluctuations, significant correlations between $K^+$ and $K^-$ are induced by the decay processes. For the ratio 
$\bar{\Lambda}/\Lambda$ we include in addition to the final particle numbers of $\Lambda$ and $\bar{\Lambda}$ also the contributions 
from weak $\Sigma^0$ and $\bar{\Sigma}^0$ decays. Those contributions were not corrected in the experimental 
analysis~\cite{Adams:2006ke,Aggarwal:2010ig,Zhao:2013yza}, while in the HRG model $\Sigma^0$ is considered a stable particle. 

%%%%%%%%%%%%%%%%%%%%%%%%%%%%%%%%%%%%%%%%%%%%%%%%%%%%%%%%%%%%%%%%%%%%%%%%%%%
\begin{figure}[t]
\centering
\includegraphics[width=0.48\textwidth]{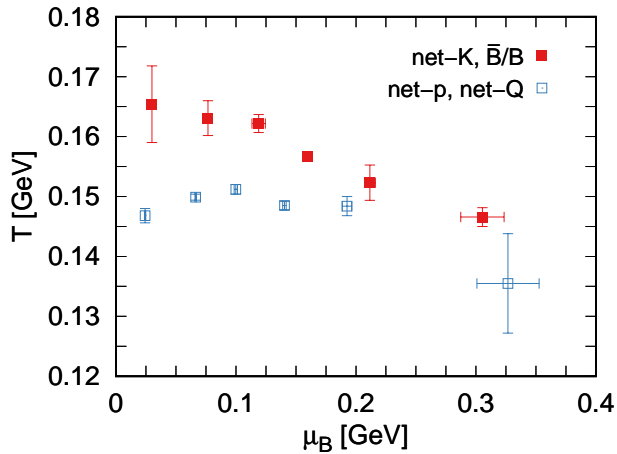}
\caption[]{\label{fig:fig2New}
(Color online) Chemical freeze-out conditions for $T$ and $\mu_B$ determined for the beam energies 
$\sqrt{s}=200, 62.4, 39, 27, 19.6, 11.5$~GeV (from left to right). The red, solid squares show the 
results determined in this work from the combined analysis of lowest-order net-Kaon fluctuations, 
$\sigma_K^2/M_K$, and $\bar{B}/B$ yield ratios. The error bars correspond to the bands in the fits shown 
in Fig.~\ref{fig:fig1New}. The blue, open squares depict the freeze-out conditions reported 
in~\cite{Alba:2014eba} which were determined from a combined analysis of lowest-order net-proton and 
net-electric charge fluctuations.}
\end{figure}
%%%%%%%%%%%%%%%%%%%%%%%%%%%%%%%%%%%%%%%%%%%%%%%%%%%%%%%%%%%%%%%%%%%%%%%%%%%
The optimal fits for $\bar{B}/B$, where the means are obtained via Eq.~(\ref{equ:mean}), and $\sigma_K^2/M_K$ including the reported 
error bars are shown by the colored bands in Fig.~\ref{fig:fig1New}. The corresponding freeze-out conditions for $T$ and $\mu_B$ are 
shown in Fig.~\ref{fig:fig2New} by the red, solid squares. For comparison, we contrast the freeze-out conditions~\cite{Alba:2014eba} 
deduced from an analysis of lowest-order net-proton and net-electric charge fluctuations which are shown by the blue, open squares. 
Similar to~\cite{Bellwied:2018tkc}, we observe in particular for large $\sqrt{s}$ a visible enhancement of the chemical freeze-out 
temperature compared to~\cite{Alba:2014eba} while for smaller $\sqrt{s}$ the two different results approach each other. Moreover, as 
in~\cite{Chatterjee:2013yga}, which bases its analysis entirely on hadronic yields distinguishing strange from non-strange hadrons, 
we find a visible but less pronounced increase in $\mu_B$. The determination of the freeze-out temperature is sensitively influenced 
by the net-Kaon fluctuation data as was already discussed in~\cite{Alba:2015iva}. Here, our use of the heavier (anti-)baryons adds 
sensitivity to $T$, while the lighter (anti-)baryons influence stronger the determination of the $\mu_B$-dependence. 

\begin{figure}[t]
\centering
\includegraphics[width=0.48\textwidth]{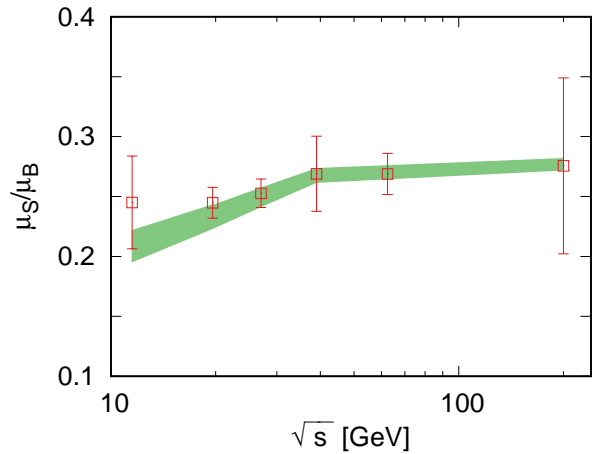}
\caption[]{\label{fig:fig3New}
(Color online) Ratio $\mu_S/\mu_B$ of strangeness over baryon chemical potential as a function of $\sqrt{s}$ 
(red, open squares) for the freeze-out conditions shown in Fig.~\ref{fig:fig2New}. The error bars account for the bands 
in the fits shown in Fig.~\ref{fig:fig1New}. Using our determined central values for $T$ and $\mu_B$ we 
show, in comparison, the corresponding lattice QCD ratio $\mu_S/\mu_B=s_1(T)+s_3(T)\mu_B^2$ (colored band), 
where the bandwidth is deduced from the data on $s_1(T)$ and $s_3(T)/s_1(T)$ and their errors reported 
in~\cite{Borsanyi:2013hza}.}
\end{figure}

The electric charge chemical potential $\mu_Q$ is negative and negligibly small compared to $\mu_B$. In contrast, the 
strangeness chemical potential $\mu_S$ constitutes a non-negligible fraction of $\mu_B$ according to the condition of 
strangeness neutrality as was stressed in previous lattice QCD studies~\cite{Borsanyi:2013hza,Bazavov:2012vg}. In 
Fig.~\ref{fig:fig3New}, we depict our results for the ratio $\mu_S/\mu_B$ (red, open squares) for the freeze-out conditions shown 
in Fig.~\ref{fig:fig2New}. For comparison, we also show the lattice QCD result $\mu_S/\mu_B=s_1(T)+s_3(T)\mu_B^2$ (colored band) 
evaluated for the central values of $T$ and $\mu_B$ determined in this work using the data on $s_1(T)$ and $s_3(T)/s_1(T)$ 
published in~\cite{Borsanyi:2013hza}. The bandwidth results from the error bars reported there. 
%%%%%%%%%%%%%%%%%%%%%%%%%%%%%%%%%%%%%%%%%%%%%%%%%%%%%%%%%%%%%%%%%%%%%%%%%%%
%%%%%%%%%%%%%%%%%%%%%%%%%%%%%%%%%%%%%%%%%%%%%%%%%%%%%%%%%%%%%%%%%%%%%%%%%%%

Finally, the importance of the fluctuation contributions due to the probabilistic character of resonance decays can be tested 
explicitly in our framework. Leaving these contributions out, the variance $\sigma_K^2$ is given by Eq.~(\ref{equ:varianceNoProb}), 
while the mean remains unaffected. Performing the analysis of the experimental data again with this set-up, we find a reduction of 
the determined freeze-out temperature by about 5\%, while the obtained $\mu_B$ is up to 18\% underestimated. This highlights that, 
indeed, the probabilistic nature of the decay processes influences our final results in a non-negligible way. 

%%%%%%%%%%%%%%%%%%%%%%%%%%%%%%%%%%%%%%%%%%%%%%%%%%%%%%%%%%%%%%%%%%%%%%%%%%%
\section{Conclusions and outlook \label{sec:4}}
%%%%%%%%%%%%%%%%%%%%%%%%%%%%%%%%%%%%%%%%%%%%%%%%%%%%%%%%%%%%%%%%%%%%%%%%%%%

In this work we determined chemical freeze-out conditions by analyzing strangeness observables from RHIC measurements at different 
beam energies. In our analysis we studied both the lowest-order net-Kaon fluctuations in~\cite{Adamczyk:2017wsl} and strange 
anti-baryon over baryon yield ratios from~\cite{Adams:2006ke,Aggarwal:2010ig,Zhao:2013yza} within a Hadron Resonance Gas model 
framework. From the combined optimized fit, the freeze-out temperature as well as the chemical potentials associated with the 
conserved charges of QCD were inferred. We find that the obtained freeze-out temperature is significantly enhanced at large 
$\sqrt{s}$ compared to the results of a previous study~\cite{Alba:2014eba} which used data on net-proton and net-electric charge 
fluctuations instead. The baryon chemical potential is also visibly enhanced compared to~\cite{Alba:2014eba} except for the 
smallest $\sqrt{s}$. For smaller $\sqrt{s}$, the freeze-out conditions from both approaches start to converge. 

The results presented here are in qualitative agreement with the findings of a recent study~\cite{Bellwied:2018tkc} which used the 
net-Kaon fluctuation data supplemented by information from lattice QCD. Our complimentary study shows that also measured yield 
ratios of strange anti-baryons over baryons can be described by similar freeze-out conditions. For a precise determination of the 
latter, the correct implementation of resonance decay contributions plays an important role. We find that resonance decays lead, 
for example, to significant correlations between $K^+$ and $K^-$ reducing the lowest-order fluctuation measure 
$\sigma_K^2/M_K$ of the net-Kaon number by about 15\% from its Skellam limit for the same thermal parameters. 

The freeze-out conditions presented in Figs.~\ref{fig:fig2New} and~\ref{fig:fig3New} are, of course, subject to both the quality 
of the analyzed data and limitations in the applied framework. The data on $\bar{B}/B$-ratios from the Beam Energy 
Scan~\cite{Zhao:2013yza} are still preliminary and therefore quantitative changes in the determined freeze-out conditions at these 
$\sqrt{s}$ can be expected for future published data, especially in terms of the considered error bars. Nevertheless, we expect 
these changes to be small in our combined analysis of net-Kaon fluctuations and $\bar{B}/B$-ratios. 

The shown errors in our results base entirely on the errors reported for the analyzed data. Additional uncertainties, stemming 
from limitations in our framework, have not been included. Different aspects that could lead to a refined analysis are, for 
example:

\noindent
(i) If the pion density in the fireball is large and the duration of the late hadronic stage long enough, the regeneration and 
subsequent decay of $K^*$ resonances provides an additional source of fluctuations in the net-Kaon number. The impact of even 
only a partial isospin randomization~\cite{Kitazawa:2011wh,Kitazawa:2012at} is to bring a distribution closer to the corresponding 
Skellam limit as we discussed in~\cite{Nahrgang:2014fza} in the case of net-proton fluctuations. To describe the 
data~\cite{Adamczyk:2017wsl} on net-Kaon fluctuations in this case would imply an enhancement of the chemical freeze-out conditions. 
Quantitative predictions of this effect would depend on the made model assumptions. 

\noindent
(ii) Exact global charge conservation realized on an event-by-event basis can cause large effects on the fluctuations. This 
effect was discussed, for example, in~\cite{Schuster:2009jv,Bzdak:2012an,Rustamov:2017lio} suggesting a future study in a canonical 
ensemble formulation. 

Our work provides a baseline which neglects dynamical as well as critical fluctuation effects on the net-Kaon number 
fluctuations. First estimates for the impact of critical fluctuations on these will be reported elsewhere. 

%%%%%%%%%%%%%%%%%%%%%%%%%%%%%%%%%%%%%%%%%%%%%%%%%%%%%%%%%%%%%%%%%%%%%%%%%%%
\section*{Acknowledgments}
%%%%%%%%%%%%%%%%%%%%%%%%%%%%%%%%%%%%%%%%%%%%%%%%%%%%%%%%%%%%%%%%%%%%%%%%%%%

The work of M. Bluhm is funded by the European Union’s Horizon 2020 research and innovation program under the Marie Sk\l{}odowska 
Curie grant agreement No 665778 via the National Science Center, Poland, under grant Polonez UMO-2016/21/P/ST2/04035. M. Nahrgang 
acknowledges the support of the program ``Etoiles montantes en Pays de la Loire 2017''. This research was supported in part by the 
ExtreMe Matter Institute (EMMI) at the GSI Helmholtzzentrum f\"ur Schwerionenforschung, Darmstadt, Germany. The authors acknowledge 
fruitful discussions within the framework of the BEST Topical Collaboration. The authors thank P.~Alba, R.~Bellwied, V.~Mantovani 
Sarti and C.~Ratti for discussions at the early stages of this work and X.~Luo for providing the data in~\cite{Adamczyk:2017wsl}. 

%%%%%%%%%%%%%%%%%%%%%%%%%%%%%%%%%%%%%%%%%%%%%%%%%%%%%%%%%%%%%%%%%%%%%%%%%%%

\end{document}